\documentclass[ amsmath,amssymb,aps,prb,reprint,superscriptaddress,showpacs]{revtex4-1}

\usepackage{graphicx}
\usepackage{dcolumn}
\usepackage{bm}
\usepackage{amsmath}

\usepackage{xcolor}

\begin{document}

\title{Quantum criticality in quasi-binary compounds of iron-based superconductors}

\author{Y.A.~Ovchenkov}
\email[]{ovtchenkov@mig.phys.msu.ru}
\affiliation{Faculty of Physics, M.V. Lomonosov Moscow State University, Moscow, 119991, Russia}
\affiliation{MIREA - Russian Technological University, Moscow, 119454, Russia }

\author{D.A.~Chareev}
\affiliation{Institute of Experimental Mineralogy, RAS, Chernogolovka, 123456, Russia}
\affiliation{Ural Federal University, Ekaterinburg, 620002, Russia }
\affiliation{Institute of Geology and Petroleum Technologies, Kazan Federal University, Kazan, 420008, Russia }
\author{A.A.~Gippius}
\affiliation{Faculty of Physics, M.V. Lomonosov Moscow State University, Moscow, 119991, Russia}
\affiliation{P.N. Lebedev Physical Institute of the Russian Academy of Science, Moscow, 199991, Russia}
\author{D.E.~Presnov}
\affiliation{Faculty of Physics, M.V. Lomonosov Moscow State University, Moscow, 119991, Russia}
\affiliation{Skobeltsyn Institute of Nuclear Physics, Moscow, 119991, Russia}
\author{I.G.~Puzanova}
\affiliation{Institute of Experimental Mineralogy, RAS, Chernogolovka, 123456, Russia}
\affiliation{National University of Science and Technology `MISiS', Moscow, 119049, Russia }
\author{ A.V.~Tkachev}
\affiliation{P.N. Lebedev Physical Institute of the Russian Academy of Science, Moscow, 199991, Russia}

\author{O.S.~Volkova}
\affiliation{National University of Science and Technology `MISiS', Moscow, 119049, Russia }
\author{S.V.~Zhurenko}
\affiliation{P.N. Lebedev Physical Institute of the Russian Academy of Science, Moscow, 199991, Russia}
\author{A.N.~Vasiliev}
\affiliation{Faculty of Physics, M.V. Lomonosov Moscow State University, Moscow, 119991, Russia}
\affiliation{National University of Science and Technology `MISiS', Moscow, 119049, Russia }


\date{\today}
%

\begin{abstract}

In this work, we present the studies of structural phase transitions in Fe(Se,Te) crystals in the range of about 30\% selenium substitution by tellurium. We found a significant change in the properties of the ordered state of these compositions compared to the case of pure FeSe. The resistivity at low temperatures for the studied Fe(Se,Te) is proportional to the square of the temperature while for pure FeSe below the structural transition it depends almost linearly on temperature. The NMR data show a noticeable line broadening below the structural transition and an anomaly in the temperature dependence of the relaxation rate in the tellurium-substituted compounds, which was not observed in the pure FeSe. This reveals in quasi-binary compounds of iron-based superconductors a region of quantum criticality similar to that which exists when the nematicity of FeSe is suppressed under pressure and which precedes the emergence of high-temperature superconductivity in FeSe under hydrostatic pressure.

\end{abstract}


\maketitle
%
%
\thispagestyle{empty}

\section{Introduction}

Superconductivity in the Iron-based superconductors (IBS) \cite{kamihara2008iron} arises in the plane formed by iron atoms. In most of these compounds, at the Fermi level there are mainly electrons of various $d-$orbitals  \cite{Johnston2010, PhysRevB.86.134520, Bohmer2018}. The energy of the $d-$electron states is determined by direct $d-d$ interaction in the plane containing the iron atoms. This interaction is strong, resulting in a large dispersion of $d$ states. The symmetry of the local environment of iron atoms also plays an important role, which is confirmed by nematic orderings.

The simplest compounds of the IBS are compositions Fe(Se,Te,S). These quasi-binary materials provide a unique opportunity to study the electronic properties of the basic building block of all IBS.

Pure FeSe is particularly fascinating among the many Fe(Se,Te,S) compositions. Firstly, this compound undergoes a structural transition without magnetic ordering. Secondly, the electronic properties of FeSe are very sensitive to different types of deformations, which means a general electronic instability possibly related to the origin of superconductivity. In particular, the tensoresistive effect exhibits extremely high values \cite{chu2012} and hydrostatic pressure can lead to an increase in the critical temperature of FeSe up to 35~K  \cite{Terashima2015,Miyoshi2014,Kothapalli2016}.

The increase in the critical temperature from 8-9~K to 35~K in structurally perfect material provides a convenient playground for identifying the necessary ingredients for high-temperature superconductivity. This increase in $T_{c}$ occurs after suppression of nematicity. At low pressure, the suppression of the structural transition in FeSe is accompanied by a decrease in $T_{c}$. Further, close to complete suppression of the structural transition and near the local minimum of $T_{c}$, magnetic ordering appears. After this, a further increase in pressure significantly increases $T_{c}$. Thus, a change in the low-temperature ground state may be directly related to the transition to the high-temperature superconductivity.

Substitution in FeSe of selenium to sulfur or tellurium also results in suppression of the structural transition temperature. For sulfur, chemical compression can be expected to be similar to the application of external pressure. For this reason, the properties of the FeSe${}_{1-x}$S${}_{x}$ series have been intensively studied in recent years. It was found that sulfur substitution at approximately $x=0.19$ is sufficient to completely suppress the structural transition \cite{Ovch_JLTP, Hosoi2016}, and this quantum critical point has attracted much attention. Later it became clear that the chemical pressure caused by selenium-sulfur substitution in FeSe${}_{1-x}$S${}_{x}$ did not have the same impact on superconducting properties as the hydrostatic pressure in FeSe. Even for FeSe${}_{1-x}$S${}_{x}$ with $0.2\le{}x\le1$ that have been obtained by the hydrothermal method \cite{PhysRevB.103.144501,wang2022suppression}, the temperature of the superconducting transition remains below 10~K. It was also demonstrated that disorder suppresses nematicity in FeSe and causes a slight increase in $T_{c}$ \cite{PhysRevB.94.064521}, which is similar to the behavior of FeSe${}_{1-x}$S${}_{x}$ at low $x$.

Recent progress in the synthesis of high-quality FeSe${}_{1-x}$Te${}_{x}$ crystals for $x\le0.5$ has contributed to the discovery of a number of interesting features of their electronic properties. \cite{Zhang182, ovchenkov2019nematic, ovchenkov2020multiband, PhysRevB.100.224516, ovchenkov2023crossover, huang2022plot}. The synthesis of high-quality crystals allowed to study in more detail the quantum criticality associated with the suppression of nematicity in FeSe${}_{1-x}$Te${}_{x}$ at low $x$. In particular, it was found that in FeSe${}_{1-x}$Te${}_{x}$ at $x\approx{}0.2-0.3$ the shape of the anomaly in the temperature dependence of resistance $R(T)$ changes in the same way as for FeSe under pressure  \cite{PhysRevB.100.224516, ovchenkov2023crossover}. This suggests that the ground state of these compounds could be magnetic.

In this paper, we report the properties of Fe(Se,Te,S) compounds that exhibit $R(T)$ anomalies which indicate a possibility of magnetic ordering. For these samples, we studied in detail their macroscopic magnetic properties, heat capacity, and NMR spectra on selenium nuclei. Our results refine the phase diagram of FeSe${}_{1-x}$Te${}_{x}$ confirming the change in the type of structural ordering at  $x\approx{}0.3$.


\begin{table*}[ht]
  \caption{\label{tbl:T1} Temperatures of phase transitions and Debye temperature of studied samples. The superconducting transition temperature $T_{c}$ is determined from ZFC measurements of magnetic susceptibility. Temperature of the structural transition $T_{S}$ is determined from the anomaly on $dR/dT(T)$ curves. The Debye temperature  $\theta_{D}$ is determined from the specific heat in the temperature range above the structural transition}
  \begin{tabular}{ccccc}
\hline
    Sample & Formula & $T_{c}$ (K) &  $T_{S}$ (K) &  $\theta_{D}$ (K)  \\
\hline
    FeSe  & FeSe  & 8.9  & 94 & 250  \\
    S-Te1 & FeSe${}_{0.67}$Te${}_{0.33}$  & 6.8  & 40 & 231  \\
    S-Te2 & FeSe${}_{0.68}$Te${}_{0.32}$ & 6.9  & 45 & 227  \\
    S-Te3 & FeSe${}_{0.675}$Te${}_{0.3}$S${}_{0.025}$ & 6.0  & 43 & 237 \\
\hline
  \end{tabular}
\end{table*}


\section{Experimental details}

Quasy-binary iron chalcogenids were grown using recrystallization in salt melts with a constant temperature gradient along the quartz ampule \cite{CrystEngComm12.1989, chareev2016general, chareev2016synthesis}.
Single-crystalline samples FeSe$_{0.67}$Te$_{0.33}$ (hereinafter named sample S-Te1),  FeSe$_{0.68}$Te$_{0.32}$ (sample S-Te2) and  FeSe$_{0.675}$Te$_{0.3}$S$_{0.025}$ (sample S-Te3) were grown and studied. For the purpose of direct comparison, we also synthesized the reference unsubstituted FeSe crystals. The chemical composition of the crystals was determined by the EDX method by averaging data over at least 20 points on five different crystals from each set. 

NMR studies on $^{77}$Se nuclei were carried out for the S-Te3 and the unsubstituted FeSe. To conduct the NMR experiment crystals of the compositions were ground and melted in paraffin to avoid shielding currents throughout the sample. All measurements were carried out in a constant magnetic field of 5.50307~T using the standard Hahn spin echo method on an upgraded Bruker MSL spectrometer  \cite{zhurenko2021}. Due to the relatively small broadening of the NMR lines in both samples, it was possible to excite them entirely at one frequency point. Therefore, the spectra were obtained as the Fourier transform of the second half of the spin echo, and relaxation measurements were carried out at the frequency of the NMR line maximum. The rate of spin-lattice relaxation was measured by the saturation recovery method.

Measurements of the magnetic susceptibility of the crystals were carried out using Quantum Design Magnetic Property Measurement System (QD MPMS) in a field of 1 Tesla in the temperature range 2-300~K. To determine the temperatures of the superconducting transition, susceptibility measurements were also carried out using the ZFC protocol in a field of 0.005 Tesla. Transport properties were measured using the EDC option for the QD MPMS. To ensure a good electrical contact, the platinum pads 100$\times$100~$\mu{}m^{2}$ in size were sputtered into the crystals and the outgoing leads were connected using conductive silver glue. Heat capacity measurements were carried out on QD Physical Property Measurement System. The measurements were carried out on sets of crystals with a total mass of 5-10~mg.


\begin{figure*}[htbp]
\centering
  \includegraphics[scale=0.5]{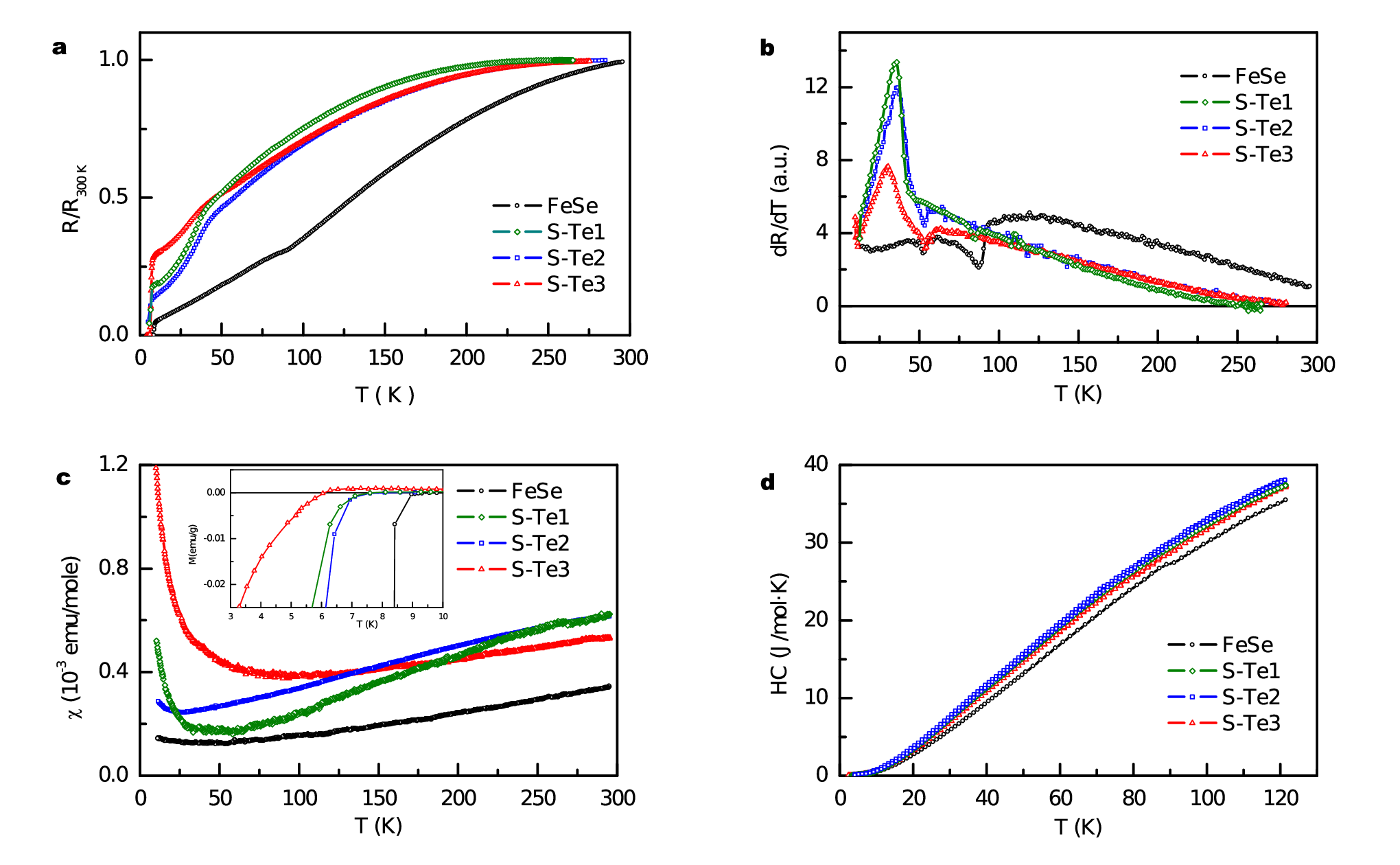}
  \caption{(a) Temperature dependence of the normalized resistance $R(T)/R(300~K)$. (b) Temperature dependence of the derivative of resistance $dR/dT$. (c) Temperature dependence of dc susceptibility  $\chi=M/H$, measured at a magnetic field of 1~T. The inset shows magnetization in 0.005 Tesla at low temperatures with zero-field-cooling (ZFC) history. (d) Temperature dependence of the heat capacity. }
  \label{fgr:fig1}
\end{figure*}

\section{Results and discussions}
\subsection{Electrical resistance, DC magnetic susceptibility and specific heat}

Fig. \ref{fgr:fig1}a) shows the temperature dependencies of the resistance of the samples, normalized to the resistance value at 300~K. We use the normalized plots because the difference in the resistivity values of the compositions at 300~K is less than the magnitude of the possible error resulting of determining the geometric parameters of the samples. The room temperature resistivities were in the range of 0.4-1.0 m$\Omega$cm for all samples. The change in the shape of $R(T)$ curves at temperatures above the structural transitions obviously indicates the transition between good and bad-metal behavior which occurs in FeSe under the substitution of selenium to tellurium. An important difference for compositions with tellurium is the shape of the anomaly on $R(T)$ at the structural transition temperature. The change is clearly visible on the $dR/dT$ curves shown in Fig. \ref{fgr:fig1}b). Similar changes of $R(T)$ curves are observed in the pure FeSe under pressure  \cite{Terashima2015,Kothapalli2016, sun2016dome}, there such a change occurs when the ground state becomes magnetic \cite{Bendele2012, Kothapalli2016}. For the Fe(Se,Te) series, a change in the anomaly on $R(T)$ appears simultaneously with the change in $R(T)$ behavior at low temperatures and suggest a transformation in the ground state \cite{ovchenkov2023crossover}. 

The temperature dependence of the static magnetic susceptibility is shown in Fig. \ref{fgr:fig1}c). All samples demonstrate a Pauli-type dependence over a wide temperature range. Structural transitions do not cause noticeable  changes in magnetic susceptibility. The increase in susceptibility at low temperatures for samples with substitution indicates the presence of a small amount of paramagnetic impurity, possibly caused by stoichiometry disorders.	Measuring susceptibility using the ZFC protocol was used to determine the temperatures of the superconducting transition, which are listed in Table \ref{tbl:T1}.

The samples with tellurium have noticeably higher specific heat values than pure FeSe (see Fig. \ref{fgr:fig1}d) ). The Debye temperatures of these compositions (Table  \ref{tbl:T1}) are higher by about 5-9 percent, which can be explained by the higher molar mass of Te. For FeSe, the $C(T)$ dependence is in good agreement with previously published data  \cite{w2008se, muratov-18} and shows a clear step at the structural transition point.  For compositions with tellurium, there are no anomalies at the temperatures of the structural transition on the $C(T)$ curves and on their temperature derivatives. This may indicate that the temperature of the structural transition  decreases due to general disorder and distortion of the local symmetry of the iron environment. Above superconducting transition temperatures, turning on a 90 kOe magnetic field has no measurable effect on the heat capacity of the samples. 

The macroscopic properties show consistent change under substitution, indicating the high quality of the grown crystals. The results of transport measurements convincingly show changes in the type of anomalies at the phase transition point which suggests the presence of a new type of quantum critical point in quasi-binary IBS. For NMR experiments, we chose the S-Te3 batch because the synthesis of this compound yielded larger and higher quality crystals.

\begin{figure}[h]
\centering
  \includegraphics[scale=0.5]{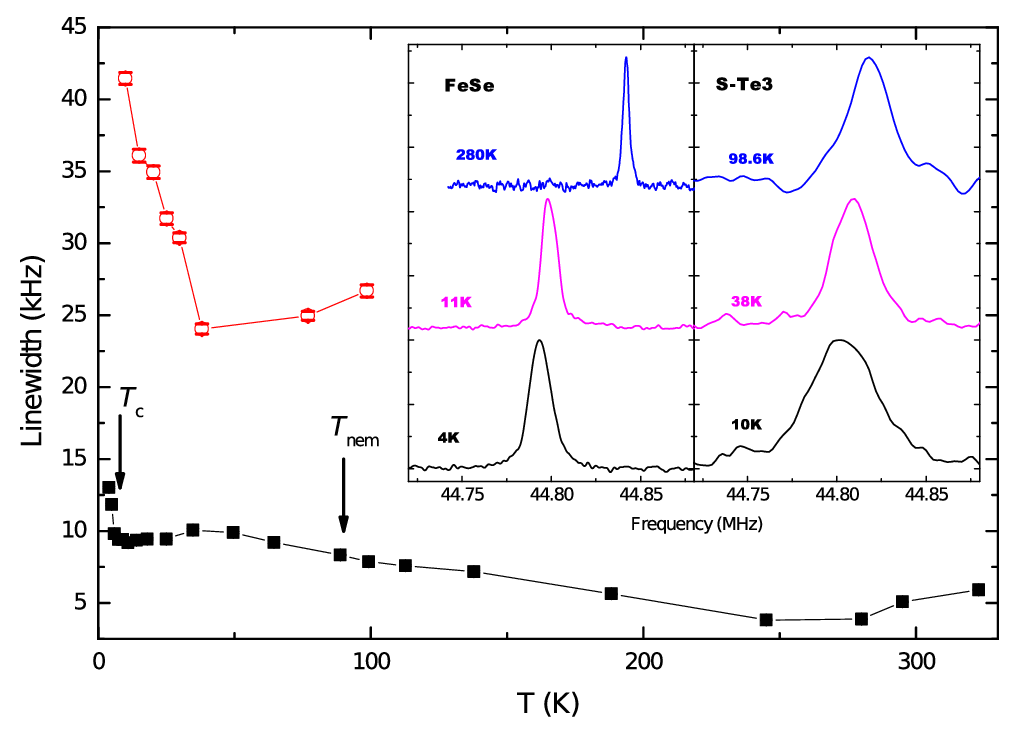}
  \caption{Temperature dependencies of the Gaussian linewidth of the $^{77}$Se NMR spectra in FeSe (black squares) and S-Te3 (red circles). Inset: the characteristic spectra of both samples at different temperatures.}
  \label{fgr:fig2}
\end{figure}

\begin{figure}[ht]
\centering
  \includegraphics[scale=0.5]{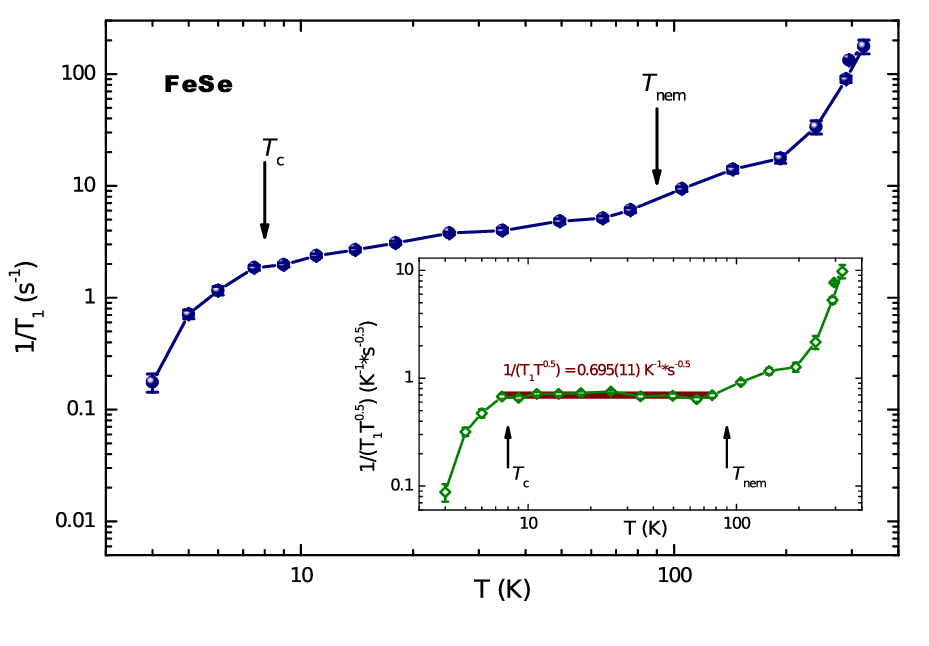}
  \caption{Temperature dependence of the spin-lattice relaxation rate 1/$T_{1}$ in the FeSe sample. Inset: temperature dependence of the reduced relaxation rate 1/($T_{1}T^{\frac{1}{2}}$) }
  \label{fgr:fig-3}
\end{figure}

\begin{figure}[ht]
\centering
  \includegraphics[scale=0.5]{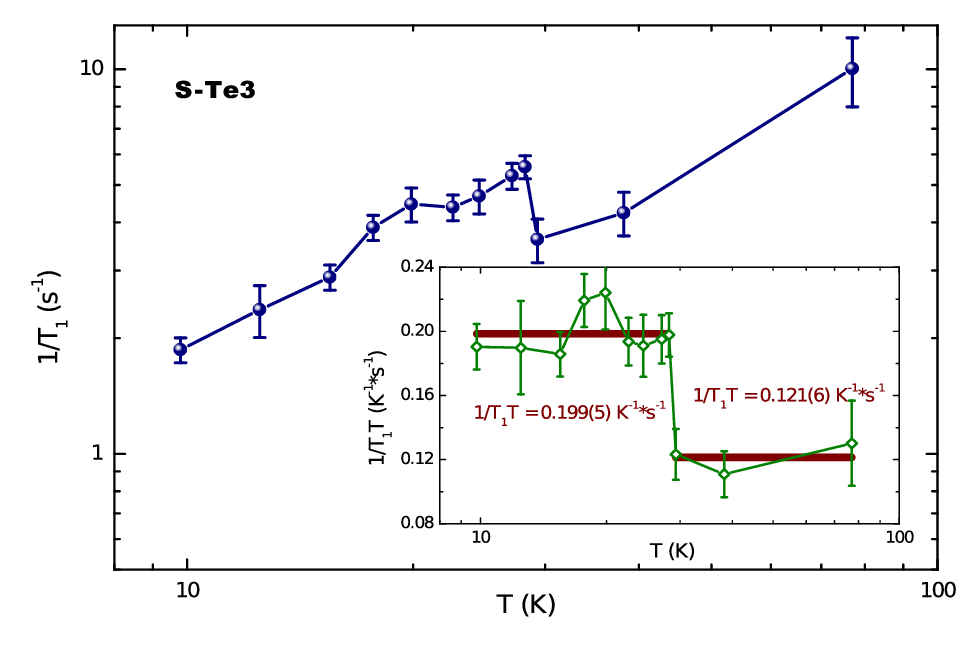}
  \caption{Temperature dependence of the spin-lattice relaxation rate 1/$T_{1}$ in the S-Te3 sample. On the inset: the temperature dependence of the reduced relaxation rate 1/$T_{1}T$.}
  \label{fgr:fig-4}
\end{figure}
\subsection{NMR}

The $^{77}$Se spectra of both powder samples show a uniformly broadened line with a shape close to Gaussian over the entire temperature range studied. Fig.\ref{fgr:fig2} demonstrates the temperature dependences of the Gaussian linewidth of the spectra together with its characteristic line shapes. The linewidth of S-Te3 is significantly higher than that for FeSe, which distinguishes this method of chemical pressure from hydrostatic pressure: the latter is not accompanied with such a pronounced broadening and anisotropy growth \cite{PhysRevLett.117.237001,PhysRevB.96.180502}.

The temperature dependence of the FeSe powder linewidth, which is determined by the shift anisotropy and the local $^{77}$Se NMR linewidth, is in good agreement with the literature data. In particular, during the transition to the superconducting state, a sharp broadening is observed, just as reported in Refs. \onlinecite{PhysRevLett.117.237001,PhysRevB.96.180502,PhysRevB.105.054514}. Further, in the temperature range from $T_{c}$ to about 50 – 60~K, the temperature-independent linewidth is consistent with constant shift anisotropy \cite{PhysRevLett.114.027001, baek2015orbital} at low values of the local linewidth \cite{PhysRevB.104.125134, PhysRevB.96.180502, PhysRevB.105.054514, PhysRevB.96.094528}. The subsequent narrowing of the line with increasing temperature is determined mainly by the decrease in anisotropy when approaching the nematic transition \cite{baek2015orbital}, which, according to Ref. \onlinecite{PhysRevLett.114.027001}, continues even with a further increase in temperature. On the other hand, in the tetragonal phase the contribution of the local linewidth increases, which, after a jump at $T_{S}$, also smoothly falls with increasing temperature (in Refs. \onlinecite{PhysRevB.96.094528, PhysRevB.104.125134, PhysRevB.96.180502} this is explained by residual nematicity), experiencing a local minimum in the region of about 200~K \cite{PhysRevB.96.180502, PhysRevB.104.125134}. This local minimum is in line with the observed minimum of powder linewidth in the range of 200 – 300~K.

The S-Te3 sample exhibits an almost constant linewidth in the range from 100 to 40~K, followed by a sharp broadening with further decrease in temperature. The breaking point ($\approx$ 40~K) coincides with the kink in the temperature dependence of the resistance ($T_{S}$ = 43~K, Fig. \ref{fgr:fig1}, Table \ref{tbl:T1}.). A possible explanation of this linewidth behavior may be related to a sharp increase in line shift anisotropy below $T_{S}$, similar to the nematic transition in the parent FeSe. It is worth noting that the linewidth increase (from 24~kHz at $T_{S}$ to 41~kHz at 10~K) is noticeably higher than that for the parent FeSe. Moreover, a pronounced brake in the temperature dependence takes place exactly at $T_{S}$, in contrast to smooth dependence for FeSe. It may point to a predominance of the contribution from anisotropy to the line broadening, rather then the local linewidth. This indicates preservation of the low-temperature nematic phase at 30\% substitution of selenium with a decrease in the transition temperature and a simultaneous increase in the degree of lattice distortion.

The temperature dependence of the spin-lattice relaxation rate 1/$T_{1}$ in FeSe (Fig. \ref{fgr:fig-3}) is also consistent with literature data. In particular, in the temperature range from $T_{c}$ to $T_{S}$, a domination of the antiferromagnetic fluctuations contribution to relaxation processes is observed, which is expressed in a rather weak growth (slower than $T^{1}$) of the relaxation rate with temperature \cite{PhysRevLett.102.177005, baek2015orbital, baek2020separate, masaki2009precise}. In the inset of Fig. \ref{fgr:fig-3} the same dependence is shown in coordinates 1/($T_{1}T^{\frac{1}{2}}$) vs $T$, which clearly shows the square root dependence of the relaxation rate in this temperature range. Indeed, in the case of weakly and almost antiferromagnetic metals, the spin-lattice relaxation rate is proportional \cite{moriya2012spin} to $T/(T-T_{N})^{\frac{1}{2}}$, which at $T\gg{}T_{N}\approx0$ gives $T^{\frac{1}{2}}$, or 1/($T_{1}T^{\frac{1}{2}}$)=const \cite{katayama1977nuclear, PhysRevB.89.104426}. The corresponding approximation (inset of Fig. \ref{fgr:fig-3}) shows good agreement with the experimental data. Below $T_{c}$ there is a sharp drop in the relaxation rate without a Hebel-Slichter peak, which is also typical for iron-containing superconductors in general and FeSe in particular \cite{PhysRevB.105.054514, PhysRevLett.102.177005, masaki2009precise, kotegawa2008evidence, PhysRevB.104.014504, shi2018pseudogap}.

The temperature dependence of the spin-lattice relaxation rate of the S-Te3 sample with 30\% substitution of tellurium for selenium (Fig. \ref{fgr:fig-4}) resembles linear behavior in the studied temperature range from 10 to 80~K. Unlike unsubstituted FeSe, in this case it is hardly possible to talk about the influence of pronounced magnetic fluctuations, because they would give slower growth  \cite{moriya2012spin}. In the 1/($T_{1}T$) vs $T$ coordinates (on the inset of Fig. \ref{fgr:fig-4}), one can clearly distinguish two horizontal regions 1/($T_{1}T$)=const, corresponding to the Korringa’s law for relaxation through conduction electrons. It is worth noting that superconductivity still takes place in this compound although partially suppressed ($T_{c}$ = 6.0~K instead of 8.9~K for parent FeSe). This result points to a rather rare case of the superconductivity preservation in iron-based superconductor without intensive magnetic fluctuations.
It resembles other tellurium intermediate substitutions FeSe$_{1-x}$Te$_{x}$ with $x = 0.5$ \cite{shimizu2009pressure}, 0.58 \cite{PhysRevB.82.140508} and 0.6 \cite{hara2011se} following Korringa’s law and contrasts with specific low-temperature magnetic enhancement of 1/($T_{1}T$) in FeSe and Te-enriched compounds FeSe$_{0.33}$Te$_{0.67}$ \cite{PhysRevB.82.064506} and FeSe$_{0.2}$Te$_{0.8}$ \cite{hara2011se}.
 Similar situation was also observed in 122 iron selenide K$_{y}$Fe$_{2-x}$Se$_{2}$ \cite{PhysRevB.83.104508, PhysRevLett.106.197001} and some other iron-based superconductors \cite{ma2013review}. This indicates that magnetic fluctuations are not the leading, or at least the only, mechanism driving to superconductivity in iron layered systems.

The observed horizontal regions on the 1/($T_{1}T$) vs $T$ plot are separated by a jump in the vicinity of $\approx$ 30~K, close to the maximum resistivity derivative $dR/dT$ (see  Fig. \ref{fgr:fig1}). When passing through a jump, the 1/($T_{1}T$) value increases by approximately 1.65 times with decreasing temperature. Such a change may be explained by two factors:

 (i) an increase in the density of states of conducting electrons at the Fermi level $N(E_{F})$. The Korringa’s law for relaxation through conduction electrons is expressed as \cite{slichter2013principles}:
 \begin{eqnarray}
 \frac{1}{T_{1}T}=\pi{}\hbar^{3}\gamma_{e}^{2}\gamma_{n}^{2}A_{hf}^{2}N^{2}(E_{F})k_{B}
\end{eqnarray}
where $\gamma_{e,n}$ are the electron and nuclear gyromagnetic ratios, $A_{hf}$ is the transferred hyperfine coupling. Therefore, the 1/($T_{1}T$) jump in Fig. \ref{fgr:fig-4} may be associated with the electron density decrease of $\sqrt{1.65}=1.28$  times above 30~K. 

(ii) Switching on the weak antiferromagnetic electron-electron interaction, which causes an enhancement of the so-called Korringa ratio $K$. The latter can be introduced by the ratio \cite{moriya1963effect, PhysRev.175.373}:
 \begin{eqnarray}
 \frac{1}{T_{1}T}=KK_{S}^{2}(\frac{\gamma_{n}}{\gamma_{e}})^{2}\frac{4\pi{}k_{B}}{\hbar}
\end{eqnarray}
where $K_{S}$ is the spin shift of the NMR line, which does not change, since the observed total shift remains almost constant ($\approx 0.3\%$) in the vicinity of this relaxation jump. $K=1$ corresponds to a non-interacting Fermi gas, while its increase (and, accordingly, the  1/($T_{1}T$) increase in S-Te3 below 30~K) may be associated with a weak antiferromagnetic correlation. Indeed, some previous studies point to the presence of the latter factor in  K$_{y}$Fe$_{2-x}$Se$_{2}$  \cite{PhysRevB.83.104508, PhysRevLett.106.197001} and FeSe$_{1-x}$S$_{x}$  \cite{PhysRevB.101.180503, PhysRevB.107.134507} and even compare it with the Fermi liquid regime. Such a comparison looks reasonable also in our case, since the resistivity temperature dependence clearly demonstrates the prevailing $\sim{}T^{2}$ component up to 30~K (Fig. \ref{fgr:fig1}a), linear growth of derivative on  Fig. \ref{fgr:fig1}b) ), which is typically considered as a Fermi liquid feature. Then the $dR/dT$ maximum can be considered as a transition to the non-Fermi liquid regime at higher temperatures, which is consistent with the NMR data.


\section{Conclusion}

Magnetic transport and thermodynamic studies of FeSe${}_{1-x}$Te${}_{x}$ show that the properties of these compositions below the structural transition change significantly in the range of tellurium concentrations of about 20-30\%. This is evidenced both by the change in the $R(T)$ power-law exponent and by the shape of the anomaly at the transition point. Properties above the temperature of structural transitions in this composition range change relatively smoothly.

From the NMR data presented, it is clear that the Te-substituted sample exhibits a number of significant differences from the pure FeSe. Firstly, it is a significant broadening of the spectral line below the structural transition, which may be a consequence of anisotropy change at the nematic transition. Secondly, there is a pronounced jump in relaxation rate slightly below the point of structural transition reflecting substantial changes in electronic structure. This behavior indicates a change in the type of nematic state with Te substitution.

The present result refines the phase diagram of FeSe${}_{1-x}$Te${}_{x}$ at low $x$, making it similar to the phase diagram of FeSe under pressure. It is interesting that in both cases, the maximum critical temperature is achieved near the quantum critical point that separates different types of nematic order although for FeSe${}_{1-x}$Te${}_{x}$ the maximum is significantly lower.

\section{Acknowledgments}
This work was supported in part by the Ministry of Science and Higher Education of the Russian Federation, grant No. 075–15-2021–1353.
 Authors acknowledge the use of the research infrastructure of the “Educational and Methodical Center of Lithography and Microscopy” of Lomonosov Moscow State University.
 OSV and DAC acknowledge Russian Science Foundation project 22-42-08002 for support of the synthesis of studied compounds. DAC acknowledges the support of the Kazan Federal University and Ural Federal University Strategic Academic Leadership Program (PRIORITY-2030)



\bibliographystyle{apsrev} 
 \bibliography{FeSeTe_2023}



\end{document}